\documentclass[12pt]{article}
\usepackage[utf8]{inputenc}
\usepackage[T1]{fontenc}
\usepackage{graphicx}
\usepackage{epstopdf} 
\usepackage{color}
\input{epsf}
\usepackage{amsmath}
\newcommand{\be}{\begin{equation}}
\newcommand{\ee}{\end{equation}}
\newcommand{\bear}{\begin{eqnarray}}
\newcommand{\ear}{\end{eqnarray}}

\begin{document}
\title{Force and torque of a string on a pulley}
\author{Thiago R. de Oliveira and Nivaldo A. Lemos  \\
\small
{\it Instituto de F\'{\i}sica - Universidade Federal Fluminense}\\
\small
{\it Av. Litor\^anea, S/N, Boa Viagem, Niter\'oi,
CEP.24210-340, Rio de Janeiro - Brazil}\\
\small
{\it  tro@if.uff.br, nivaldo@if.uff.br}}

\date{\today}

\maketitle

\begin{abstract}

Every university introductory physics course considers the problem of Atwood's machine taking 
into account the mass of the pulley. In the usual treatment the tensions at the two ends of 
the string are offhandedly taken to act on the pulley and be responsible for its rotation. However such 
a free-body diagram of the forces on the pulley is not {\it a priori} justified, inducing 
students to construct wrong hypotheses such as  that the string transfers its tension to the pulley or that  some symmetry is in operation. We reexamine this problem by integrating the contact forces between each element of the string and the pulley and show that although the pulley does behave as if the tensions were acting on it, this  comes only as the end result of a  detailed analysis. We also address the question of how much friction is needed to prevent the string from slipping over the pulley. Finally,  we deal with the case in which  the string is on the verge of sliding and show that this will never happen unless certain conditions are met by the  coefficient of friction and the masses involved.

\end{abstract}

\section{Introduction}

\begin{figure}[h!] 
\centering
\includegraphics[width=0.2\linewidth]{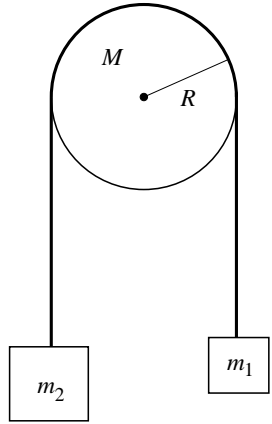}
\caption[Circle]{Two masses attached to a massless string that goes around a massive pulley.}
\label{masses-pulley}
\end{figure}

A crucial step in solving a problem in mechanics by applying Newton's laws to interacting  bodies 
is to identify the individual forces that act on each object. We dare  say that the physics ends
there and the rest is only manipulation of equations. Although this may be  too strong a 
statement, it has been recognized in recent years that many students have difficulties with 
this step and that textbooks and instructors should give more attention to the correct 
identification of the forces on each body \cite{Knight}. One type of problem in which most,
if not all, textbooks fail to correctly identify the forces on each object are
the ones containing  strings and  massive pulleys. These problems are  treated  in any university 
elementary physics course that addresses the rotational dynamics of  rigid bodies about 
a fixed axis. 
A staple problem is  Atwood's machine \cite{Greenslade} with a pulley whose mass $M$ is not 
negligible in comparison with $m_1$ and $m_2$, depicted in Fig. \ref{masses-pulley}. As  usual, 
we assume that  the string is inextensible, its mass is negligible and it does not slide on the pulley, which requires enough static friction between the string and the pulley.
On the other hand, we assume that the pulley is mounted on a frictionless axle.

\begin{figure}[h!]
\centering
\includegraphics[width=0.3\linewidth]{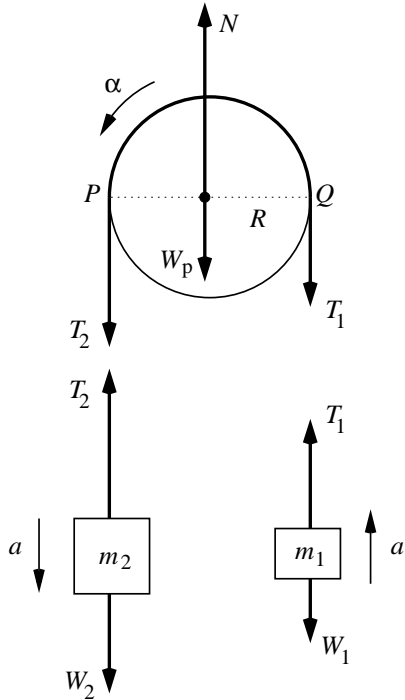}
\caption[Circle]{Standard free-body diagram of forces on the masses and the pulley.}
\label{forces-masses-pulley}
\end{figure}

For the sake of definiteness let us assume that $m_2 > m_1$. The standard free-body diagram of forces on the hanging masses and the pulley is shown in Fig. \ref{forces-masses-pulley}.  Since $W_1=m_1g$ and $W_2=m_2g$, Newton's second law  applied to the masses gives 
\begin{subequations}
\label{Newton-2nd-law-masses}
\begin{align}
\label{Newton-2nd-law-mass1}
T_1 - m_1 g  & = m_1 a \, , \\
\label{Newton-2nd-law-mass2}
m_2 g - T_2 & = m_2 a \, . 
\end{align}
\end{subequations}	

As to the pulley, the standard claim \cite{Serway} is that $T_2$ and $T_1$ are forces 
exerted by the string {\it on the pulley} at the points $P$ and $Q$, respectively. 
The weight of the pulley and the sustaining force exerted by the  axle produce no torque about 
the rotation axis. Therefore, the net torque on the pulley is $\tau = T_2R - T_1R$. Applying the
so-called Newton's second law for rotational motion, $ \tau = I\alpha $,
to the pulley one has
\begin{equation}
\label{Newton-2nd-law-pulley}
(T_2 - T_1)R =I\alpha\, .
\end{equation}	

Since the string does not slide on the pulley, there holds the constraint $a= \alpha R$. With the use of this constraint and the assumption that the pulley is a homogeneous disk, whose pertinent moment of inertia is $I=MR^2/2$, equation (\ref{Newton-2nd-law-pulley}) becomes
\begin{equation}
\label{Newton-2nd-law-pulley2}
T_2 - T_1 =\frac{M}{2}  a\, .
\end{equation}	
By simply summing equations (\ref{Newton-2nd-law-mass1}),  (\ref{Newton-2nd-law-mass2}) and (\ref{Newton-2nd-law-pulley2}) one gets the acceleration, and with  a little more elementary algebra one finds the tensions. The result is
\begin{equation}
\label{acceleration-tensions}
a = \frac{m_2 -m_1}{m_1+m_2+M/2} g\, , \,\,\,\,\, T_1=\frac{2m_2+M/2}{m_1+m_2+M/2}m_1g \, , \,\,\,\,\, T_2=\frac{2m_1+M/2}{m_1+m_2+M/2}m_2g  \, .
\end{equation}	

This standard solution  to the problem of the motion of Atwood's machine with a massive pulley, as well as the solutions  to similar problems that can be found in so many textbooks, is open to a serious physical objection. The forces $T_2$ and $T_1$ are {\it not} forces on the pulley, but forces on the points $P$ and $Q$  {\it of the  string} exerted by the hanging parts of the string at each side of the pulley. This gives rise to the problem of justifying the above results obtained on the basis of  a physically unwarranted identification of the forces on 
the pulley. Note that putting directly the forces $T_2$ and $T_1$ on the pulley might
reinforce a common misconception among students that all strings do is convey forces from one  object to another.
Here the problem is even more subtle since we do consider the string as massless,  which usually 
entails that the tension is constant all along the string. 
There is also an interesting question that, as far as we can tell, is  never asked in the textbooks: What is the magnitude of the friction force that prevents the string from slipping on the pulley? 

The proper physical analysis consists in taking into account that each element of the string exerts a force on the part of the  pulley with which it is in contact \cite{Krause}. The determination of the net force and the net torque on the pulley requires an integration of the infinitesimal forces and torques exerted on the pulley by each element of the string that touches the pulley. 
Here one might be tempted to  justify the usual treatment 
by the seemingly  reasonable   conjecture that, except for the forces at points $P$ and $Q$ of the pulley,  the vector sum of all forces exerted by the the string on the other points of the pulley cancel each other owing to an apparent symmetry. This argument turns out to be wrong, and no such symmetry exists.

\section{Forces on an element of the string}
 
We follow the approach used in the analysis of the related problem of determining the effect of the friction force on a rope wrapped around a capstan \cite{Becker,Agmon}.
  
\begin{figure}[h]
\centering
\includegraphics[width=0.4\linewidth]{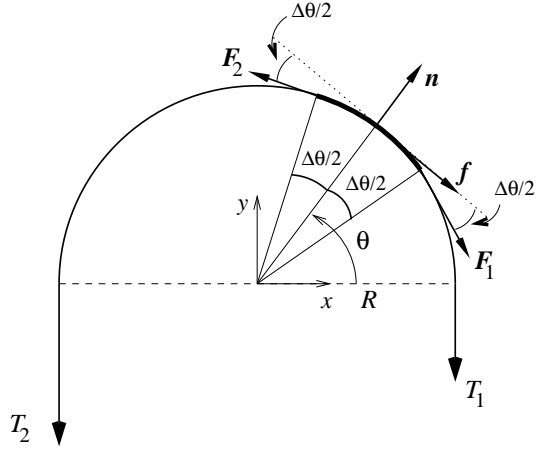}
\caption[Circle]{Forces  on a piece of the string that subtends the small angle $\Delta \theta$.}
\label{forces-string-piece}
\end{figure}
	
For the sake of definiteness we insist on the assumption that $m_2 > m_1$, which implies $T_2 > T_1$. 	Figure \ref{forces-string-piece} shows the forces  on a piece of the string that subtends the small angle $\Delta \theta$. ${\bf F}_1$ and ${\bf F}_2$ are the tensions at the ends of the string element; $\bf f$ and $\bf n$  are, respectively,  the tangential (friction) and normal forces exerted by the pulley.
Since the string is massless, Newton's second law entails that the vector sum of the forces shown in Figure \ref{forces-string-piece} is zero:
\begin{equation}
\label{total-force-zero}
{\bf F}_1 + {\bf F}_2 + {\bf f} + {\bf n} =0 \, .
\end{equation}	
Let
 \begin{equation}
\label{unit-vectors}
{\hat {\bf r}}=  \cos \theta \, {\hat {\bf x}} + \sin \theta \, {\hat {\bf y}}\, , \,\,\,\,\,\,\,\,\,\, {\hat {\mbox{\boldmath $\theta$}}}= -\sin \theta \, {\hat {\bf x}} + \cos \theta \, {\hat {\bf y}} 
\end{equation}
 respectively be the outward normal and tangential (oriented toward increasing $\theta$) unit 
 vectors  at the point of the pulley with angular coordinate $\theta$.
For the the friction force and the normal force of the pulley  {\it on the piece of  string}
we write
\begin{equation}
\label{friction-normal-forces}
{\bf f} = -f R \Delta \theta \, {\hat {\mbox{\boldmath $\theta$}}}\, , \,\,\,\,\,\,\,\,\,\, {\bf n} = n R \Delta \theta \, {\hat {\bf r}} 
\end{equation}
where $f$ and $n$ are positive and have dimension of force per unit length.

The tangential and  normal components of   equation
 (\ref{total-force-zero}) are
\begin{subequations}
\label{components-total-force}
\begin{align}
\label{tangential-total-force}
F_2 \cos \bigl( \frac{\Delta \theta}{2}\bigr) - F_1 \cos \bigl( \frac{\Delta \theta}{2}\bigr) -fR\Delta \theta  & = 0 \, , \\
\label{normal-total-force}
-F_1 \sin  \bigl( \frac{\Delta \theta}{2}\bigr) - F_2 \sin \bigl( \frac{\Delta \theta}{2}\bigr) + nR\Delta \theta & = 0 \, . 
\end{align}
\end{subequations}
In  equations	(\ref{components-total-force})   we have 	
	\begin{equation}
\label{F1-and-F2}
F_2 = \vert {\bf F}_2 \vert = T \bigl(\theta + \frac{\Delta \theta}{2}\bigr), \,\,\,\,\,\,\,\,\,\,  F_1 = \vert {\bf F}_1 \vert = T \bigl(\theta - \frac{\Delta \theta}{2}\bigr) \, ,
\end{equation}	
where $T( \theta )$ is the  tension at
the point of the string with angular coordinate $\theta$ with, of course,
\begin{equation}
\label{T1-and-T2-zero-pi}
T (0) = T_1 \, , \,\,\,\,\,\,\,\,\,\,  T (\pi ) = T_2 \, .
\end{equation}
Therefore Eqs.(\ref{components-total-force}) become 
\begin{subequations}
\label{components-total-force-final}
\begin{align}
\label{tangential-total-force-final}
 T \bigl(\theta + \frac{\Delta \theta}{2}\bigr) \cos \bigl( \frac{\Delta \theta}{2}\bigr) - T \bigl(\theta - \frac{\Delta \theta}{2}\bigr) \cos \bigl( \frac{\Delta \theta}{2}\bigr) -fR\Delta \theta  & = 0 \, , \\
\label{normal-total-force-final}
-T \bigl(\theta - \frac{\Delta \theta}{2}\bigr)\sin  \bigl( \frac{\Delta \theta}{2}\bigr) - T \bigl(\theta + \frac{\Delta \theta}{2}\bigr) \sin \bigl( \frac{\Delta \theta}{2}\bigr) + nR\Delta \theta & = 0 \, . 
\end{align}
\end{subequations}

Now we divide  each of the two last equations by $\Delta \theta$ and let 
$\Delta \theta \to 0$ to obtain
\begin{equation}
\label{f}
fR = \frac{dT}{d \theta}
\end{equation}	
and also
\begin{equation}
\label{n}
nR = T ( \theta) \, .
\end{equation}		
Equation (\ref{f}) shows that friction causes the tension in the string to be variable, even though the string is massless.

\section{Force on the pulley}

We are now in a position to compute the friction force and the total force exerted by the string on the pulley.

The net friction force exerted by the  pulley {\it on the the string} is given by
\begin{equation}
\label{total-friction-on-string}
{\bf F}_f^{string} = \int_0^{\pi}(-f R \, {\hat {\mbox{\boldmath $\theta$}}}) d \theta
= {\hat {\bf x}}\int_0^{\pi}\frac{dT}{d \theta}\sin\theta d\theta -
{\hat {\bf y}}\int_0^{\pi}\frac{dT}{d \theta}
 \cos \theta d \theta
\end{equation}
where we have used (\ref{unit-vectors}), (\ref{friction-normal-forces}) and (\ref{f}).
Integrations by parts yield 
\begin{equation}
\label{integration-parts-1}
\int_0^{\pi}\frac{dT}{d \theta}\sin\theta d\theta = T ( \theta )\sin\theta\bigg\vert_0^{\pi} - \int_0^{\pi} T (\theta )\cos\theta d\theta =
- \int_0^{\pi}T (\theta )\cos\theta d\theta
\end{equation}	
and 
\begin{equation}
\label{integration-parts-2}
\int_0^{\pi}\frac{dT}{d \theta}\cos\theta d\theta = T ( \theta )\cos\theta\bigg\vert_0^{\pi} + \int_0^{\pi}T (\theta )\sin\theta d\theta =
-(T_1+T_2) + \int_0^{\pi}T (\theta )\sin\theta d\theta \, .
\end{equation}
With these results Eq. (\ref{total-friction-on-string}) becomes
\begin{equation}
\label{total-friction-on-string-final}
{\bf F}_f^{string} = (T_1+T_2) {\hat {\bf y}} - \int_0^{\pi}T (\theta )(\cos \theta \, {\hat {\bf x}} + \sin \theta \, {\hat {\bf y}} ) d\theta\, .
\end{equation}
One could think that by symmetry the $x$-component in (\ref{total-friction-on-string-final}) would be zero, but this is not true because $T(\theta)\neq T(\pi-\theta)$. We also note that the friction force on the string cannot be determined unless the tension is known as a function of $\theta$. Thus, in general, the question ``What is the magnitude of the friction force that prevents the string from slipping over the pulley?'' does not have a definite answer. As will be seen shortly, however, $T(\theta )$
can be found explicitly if the string is on the verge of sliding on the pulley. 

The net normal force exerted by the  pulley {\it on the the string} is given by
\begin{equation}
\label{total-normal-force-on-string}
{\bf F}_n^{string} = \int_0^{\pi}n R \, {\hat {\bf r}}\, d \theta
= \int_0^{\pi}T (\theta )(\cos \theta \, {\hat {\bf x}} + \sin \theta \, {\hat {\bf y}} ) d\theta\, ,
\end{equation}
where equations (\ref{unit-vectors}), (\ref{friction-normal-forces}) and  (\ref{n}) have been used. 
From (\ref{total-friction-on-string-final}) and (\ref{total-normal-force-on-string}) it follows at once that
\begin{equation}
\label{total-force-pulley-on-string}
{\bf F}_f^{string} +  {\bf F}_n^{string} - (T_1+T_2) {\hat {\bf y}} = 0 \, .
\end{equation}
Note that $\, - (T_1+T_2) {\hat {\bf y}}  \,$  is the total force on the part of the string  in contact with the pulley  exerted by the hanging pieces of the string at each side of the pulley. Thus the last result  is correct: the total force on the part of the string in contact with the pulley is zero because the string is massless.

By Newton's third law, the total force exerted  by the string {\it on the pulley} is
\begin{equation}
\label{total-force-pulley-on-string}
{\bf F}_{pulley} = - ({\bf F}_f^{string} +  {\bf F}_n^{string}) =  - (T_1+T_2) {\hat {\bf y}}  \, .
\end{equation}
Therefore, although $T_2$ and $T_1$ {\bf are not} forces on the pulley, everything  happens as if they actually were forces applied by the string  at the points $P$ and $Q$ of the pulley shown in Fig. \ref{forces-masses-pulley}, and as if the forces exerted by the string on  the  other points of the pulley cancelled each other owing to an  apparent (but nonexistent) symmetry.

\section{Torque on the pulley}

The torque exerted by the string  on an element {\it of the pulley} that subtends an angle $d\theta$ is 
\begin{equation}
\label{element-torque-on-pulley1}
d{\mbox{\boldmath $\tau$}}_{pulley} = R{\hat{\bf r}} \times (-{\bf f}) =
R{\hat{\bf r}} \times {\hat {\mbox{\boldmath $\theta$}}} fR d\theta = fR^2 {\hat {\bf z}} d\theta
\end{equation}
where (\ref{friction-normal-forces}) has been used.  By making use of (\ref{f}) we are led to
\begin{equation}
\label{element-torque-on-pulley2}
d{\mbox{\boldmath $\tau$}}_{pulley} = {\hat {\bf z}} R \frac{dT}{d\theta}  d\theta  \, .
\end{equation}
Therefore,
\begin{equation}
\label{torque-on-pulley1}
{\mbox{\boldmath $\tau$}}_{pulley} = {\hat {\bf z}} R\int_0^{\pi}\frac{dT}{d\theta}  d\theta = R(T_2-T_1) {\hat {\bf z}} \, .
\end{equation}
Once again, this is the torque on the pulley obtained by the {\it a priori} physically unwarranted assumption that  $T_2$ and $T_1$  are  forces on the pulley and that the torques applied by the string on the pulley at  points other than the points $P$ and $Q$  shown in Fig. \ref{forces-masses-pulley} cancel each other owing to an  apparent (but nonexistent) symmetry. It should be noted  that one can calculate directly the total torque on the pulley without first finding the net  frictional  and normal forces as has been done here \cite{Krause}.
 
\section{String  on the verge of sliding}

Let us suppose that the string is on the verge of sliding on the pulley. If $\mu$ is the coefficient of static friction between the string and the
pulley we have
\begin{equation}
\label{f-mu-n}
f = \mu n \, .
\end{equation}
Combining this equation with (\ref{f}) and (\ref{n})  we find
\begin{equation}
\label{diff-eqn-T}
\frac{dT}{d\theta} = \mu T \, .
\end{equation}
It follows that
\begin{equation}
\label{T-of-theta}
T({\theta}) = T_1 e^{\mu\theta}
\end{equation}
inasmuch as $T(0) = T_1$. Since $T_2 = T (\pi ) =  T_1 e^{\mu\pi}$, the friction coefficient is determined:
\begin{equation}
\label{mu-determined}
\mu = \frac{1}{\pi} \ln \Bigl(\frac{T_2}{T_1}\Bigr) \, .
\end{equation}
By the way, the exponential growth of the tension explains why, if a rope is wound several times around a capstan, it takes an enormous force to make the rope slide  on the capstan by pulling  one end  against a tiny force at the other end   \cite{Becker,Agmon}.

Now the friction force {\it on the pulley} can be explicitly computed. From (\ref{total-friction-on-string-final}) and (\ref{T-of-theta}) we have
\begin{eqnarray}
\label{total-friction-on-pulley-verge}
{\bf F}_f^{pulley} & = & -{\bf F}_f^{string} = - (T_1+T_2) {\hat {\bf y}} + \int_0^{\pi}T (\theta )(\cos \theta \, {\hat {\bf x}} + \sin \theta \, {\hat {\bf y}} ) d\theta \nonumber \\
& = & - (T_1+T_2) {\hat {\bf y}} + {\hat {\bf x}}T_1 \int_0^{\pi}e^{\mu\theta}\cos \theta d\theta +  {\hat {\bf y}}T_1 \int_0^{\pi}e^{\mu\theta}\sin \theta d\theta \, .
\end{eqnarray}
The integrals are elementary and can  also be found in any table:
\begin{eqnarray}
\label{integrals}
\int_0^{\pi}e^{\mu\theta}\cos \theta d\theta & = & \frac{e^{\mu\theta}}{1+\mu^ 2}(\sin  \theta + \mu\cos \theta )\bigg\vert_0^{\pi} = - \frac{\mu}{1+\mu^ 2}( 1+ e^{\mu\pi})\, ;\nonumber \\
\int_0^{\pi}e^{\mu\theta}\sin \theta d\theta & = & \frac{e^{\mu\theta}}{1+\mu^ 2}(\mu \sin  \theta - \cos \theta )\bigg\vert_0^{\pi} =  \frac{1}{1+\mu^ 2}( 1+ e^{\mu\pi})\, . 
\end{eqnarray}
Therefore,
\begin{equation}
\label{total-friction-on-pulley-verge-explicit}
{\bf F}_f^{pulley} = - (T_1+T_2) {\hat {\bf y}} - \frac{\mu T_1}{1+\mu^ 2}( 1+ e^{\mu\pi}) {\hat {\bf x}} + \frac{T_1}{1+\mu^ 2}( 1+ e^{\mu\pi}) {\hat {\bf y}} 
 \, .
\end{equation}
With the help of (\ref{mu-determined}) and a little algebra this last result can be cast in the following  form:
\begin{equation}
\label{total-friction-on-pulley-verge-final}
{\bf F}_f^{pulley} = -\frac{\mu }{1+\mu^ 2} (T_1+T_2) {\hat {\bf x}}  -\frac{\mu^2}{1+\mu^ 2} (T_1+T_2) {\hat {\bf y}} 
 \, .
\end{equation} 
Now we have a definite answer to our previous question:  the magnitude of the friction force that prevents slippage of the string over the pulley is 
\begin{equation}
\label{magnitude-total-friction-on-pulley-verge-final}
F_{f}^{\, pulley} = \frac{\mu}{\sqrt{1+\mu^2}}  (T_1+T_2) 
 \, .
\end{equation}

From  (\ref{acceleration-tensions}) and 
(\ref{mu-determined}) it follows that
\begin{equation}
\label{T2-over-T1}
\frac{T_2}{T_1} =\frac{(4m_1+M)m_2}{(4m_2+M)m_1} = e^{\mu \pi} \, .
\end{equation}	
Solving this equation for $M$ we find
\begin{equation}
\label{M}
M =\frac{4m_1m_2( e^{\mu \pi} - 1)}{m_2 - m_1 e^{\mu \pi}} \, .
\end{equation}
Note that if $m_2 \leq m_1  e^{\mu \pi}$ then the string will never slip relative  to the pulley no matter how large the pulley mass is.
On the other hand, solving (\ref{T2-over-T1}) for $m_2$ we get
\begin{equation}
\label{m2}
m_2 =\frac{Mm_1 e^{\mu \pi}}{M - 4m_1 (e^{\mu \pi}-1)} \, .
\end{equation}
If $M \leq 4m_1 (e^{\mu \pi}-1)$ there is no positive solution for $m_2$. Therefore, two necessary conditions for the string to be on the the verge of slipping over the pulley are
\begin{equation}
\label{ecessary-conditions}
m_2 > m_1  e^{\mu \pi}\,\,\,\,\,\,\,\,\,\, \mbox{and} \,\,\,\,\,\,\,\,\,\, M > 4m_1 (e^{\mu \pi}-1) \, .
\end{equation}
The first requirement is expected from the  force amplification effect brought about  by a  rope wrapped around 
a capstan \cite{Becker,Agmon}. It would be the only necessary condition if the pulley could not rotate or, equivalently, if its 
mass were infinite. The second requirement is not so obvious and stems from the fact that the pulley can 
freely turn on its axle.

So as to have an idea of the order of magnitude of the masses that are required for 
the string to be on the verge of sliding, let us assume that $m_1 = 1$ kg,  $m_2 = 3$ kg 
and $\mu = 0.3$. Then (\ref{M}) gives  $M \approx 43$ kg, an appreciably large mass for the pulley. 

\section{Concluding Remarks}

We have argued that the usual textbook solution to a classic problem in rotational dynamics, Atwood's machine, relies on a faulty   identification of forces on each object, since  the tensions in the hanging parts of the string are identified  as forces on the pulley. This may seem a small detail, but we believe it is an important one, since correctly identifying the forces on each of a system of  interacting bodies is a fundamental  step in solving a mechanics problem by means of  Newton's laws. Such a shortcut also reinforces a common misconception between students to the effect  that strings merely convey forces without affecting them. 

We have presented a consistent  treatment of the problem by 
considering  the normal and friction forces between each element of the string and the pulley, and have shown  that the contact force exerted  on 
the pulley by the entire segment of  the string that touches the pulley gives rise to the net force and the net torque usually assumed without a convincing justification in the standard treatment given in the textbooks.

Although  presenting the full mathematical treatment of the problem, as done here, is beyond the scope of an introductory physics 
course, we believe that attention should be called to the fact that the force on the pulley
arises from contact forces exerted by  the string and that a careful
analysis gives the conjectured result (\ref{torque-on-pulley1}) for the torque on the pulley. Another possibility is  to consider the pulley together with the segment of the string that touches it as a single object with the same moment of inertia as that of the pulley, since the string is massless.  As far as the angular acceleration of this object is 
concerned,  one is allowed to disregard the internal forces between the aforesaid segment of the string and the pulley, and now $T_1$ and $T_2$ are actually external forces responsible for the net external torque on the pulley-string segment system \cite{Krause}.

\subsection*{Acknowledgements}
TRO is supported by the National Institute for Science and Technology of Quantum Information (INCT-IQ) and also thanks the National Council for Scientific and Technological Development (CNPq), Brazil.


\newpage
 
\begin{thebibliography}{99}
\bibitem{Knight} R. D. Knight, {\it Five Easy Lessons: Strategies for Successful Physics Teaching} (Addison-Wesley, San Francisco, 2004), chap. 7.

\bibitem{Greenslade} For a historical account of this iconic system,
see   T. B. Greenslade, Jr.,  ``Atwood's machine,'' {\it Phys.
Teach.} {\bf 23},   24-28 (1985).


\bibitem{Serway} See, for example, R. A. Serway and J. W. Jewett, Jr., {\it Principles of Physics}, 4th ed. (Thomson Brooks/Cole, Belomon, CA, 2006), pp. 310-311

\bibitem{Krause} D. E. Krause and Y. Sun, ``Can a string's tension exert a torque on a pulley?'' {\it Phys. Teach.} {\bf 49}, 234-235 (2011).

\bibitem{Becker} R. A. Becker, {\it Introduction to Theoretical Mechanics}
 (McGraw-Hill, New York, 1954), pp. 45-47.   

\bibitem{Agmon} D. Agmon and P. Gluck, {\it Classical and Relativistic Mechanics} (World Scientific, Singapore, 2009), pp. 117-118.





\end{thebibliography}
\end{document}